\documentclass[aps,pra,floatfix,superscriptaddress,twocolumn]{revtex4-1} 

\usepackage{graphicx}
\usepackage{graphics}
\usepackage{amssymb}
\usepackage{amsmath}
\usepackage{epsfig}
\usepackage{latexsym}
\usepackage{color}

\newcommand{\bra}[1]{\left\langle{#1}\right\vert}
\newcommand{\ket}[1]{\left\vert{#1}\right\rangle}

\newcommand{\expect}[1]{\langle{#1}\rangle}

\newcommand{\beq}{\begin{equation}}
\newcommand{\eeq}{\end{equation}}
\newcommand{\bqa}{\begin{eqnarray}}
\newcommand{\eqa}{\end{eqnarray}}
\newcommand{\nn}{\nonumber}

\newcommand{\erf}[1]{Eq.~(\ref{#1})}
\newcommand{\dg}{^\dagger}
\newcommand{\dd}[1]{\textrm{d}{#1}}
\newcommand{\spa}{~~~~~~}

\begin{document}

\title{Reduced equations of motion for quantum systems driven by diffusive Markov processes}
\author{Mohan Sarovar}
\affiliation{Department of Scalable and Secure Systems Research, Sandia
  National Laboratories, Livermore, CA 94550} 
\author{Matthew D. Grace}
\affiliation{Department of Scalable and Secure Systems Research, Sandia
  National Laboratories, Livermore, CA 94550} 

\begin{abstract}
The expansion of a stochastic Liouville equation for the coupled
evolution of a quantum system and an Ornstein-Uhlenbeck process into a
hierarchy of coupled differential equations is a useful technique that
simplifies the simulation of stochastically-driven quantum systems. We
expand the applicability of this technique by completely characterizing
the class of diffusive Markov processes for which a useful hierarchy of
equations can be derived. The expansion of this technique enables the
examination of quantum systems driven by non-Gaussian stochastic
processes with bounded range. We present an application of this extended technique by simulating Stark-tuned F\"orster resonance transfer in Rydberg atoms with non-perturbative position fluctuations.
\end{abstract}

\maketitle

Describing the dynamics of a quantum system coupled to uncontrolled degrees of freedom has been an important problem since the inception of quantum mechanics, e.g., \cite{Pauli:1928ta}. The accurate description of such \textit{open quantum systems} is particularly vital for the design of quantum technologies, such as quantum computers. Several approximate and exact methods exist for describing the dynamics of open quantum systems, including master equations, surrogate Hamiltonians, and Monte-Carlo numerical simulations. 

In this work, we examine quantum systems driven linearly by classical
fluctuations. That is, the Hamiltonian for the system is described by
$\hat{H}_\Omega(t) = \hat{H}_0 + \Omega(t)\hat{V}$, where $\Omega(t)$ is a
time-dependent stochastic variable. This is a sub-class of a more general open quantum system where $\Omega(t)$ is an operator in the Hilbert space of an uncontrolled environment. The replacement of the operator with a scalar variable is an approximation that is valid in certain limits (e.g., in the high temperature limit of a bosonic environment). Scalar fluctuations can also describe noise in a quantum system that is controlled by an effectively classical quantity, such as a gate voltage. 

For such stochastic evolution, the dynamics of the system under a \textit{given} trace of the noise is dictated by the von-Neumann (Schr\"odinger) equation: 
\beq
\frac{\partial}{\partial t} \hat{\rho}(t | \{\Omega(t)\}) = -\frac{i}{\hbar} [\hat{H}_\Omega(t), \hat{\rho}(t | \{\Omega(t)\})].
\label{eq:rho_evol}
\eeq
Here, the notation $\hat{\rho}(t | \{\Omega(t)\})$ is used to explicitly
indicate that this density matrix is \emph{conditioned} on a particular
realization of the noise process. This equation is formally solved to yield $\hat{\rho}(t | \{\Omega(t)\}) = \hat{\mathcal{U}}(t,0) \hat{\rho}^0 \hat{\mathcal{U}}(t,0)\dg$, for initial state $\rho^0$, where
$\hat{\mathcal{U}}(t,0) = \mathcal{T} e^{-\frac{i}{\hbar}\int_0^t \hat{H}_\Omega(t)}$
with $\mathcal{T}$ being the time-ordering operator. 
We are often more interested in the unconditioned evolution of the system state, after the fluctuating quantity has been averaged:
\beq
\hat{\rho}(t) = \expect{\hat{\rho}(t | \{\Omega(t)\})}_{\{\Omega(t)\}},
\label{eq:rho_functional}
\eeq
where the angled brackets denote an expectation value over the
stochastic process up to time $t$. We will refer to differential
equations describing the evolution of this averaged quantity as
\textit{reduced equations of motion}. In this work we derive reduced equations of motion for a quantum system coupled to a wide family of stochastic processes.

We focus on stochastic processes that are time-continuous,
time-homogenous, and Markov,
which means that the evolution of the conditional probability distribution for the process evolves as \cite{Risken:1984vl, Gar-2004, Kam-2007}
\beq
\frac{\partial}{\partial t}P(\Omega, t | \Omega', t') = \Gamma_\Omega P(\Omega, t | \Omega', t'),
\label{eq:markov_generator}
\eeq
where $P(\Omega, t | \Omega', t')$ is the probability that the stochastic process takes the value $\Omega$ at time $t$ given that it took the value $\Omega'$ at time $t'$ ($t' \leq t$). $\Gamma_\Omega$ is the forward \textit{generator} of the process and is a differential operator only involving derivatives with respect to $\Omega$. The generator of evolution is in general a complex quantity that results from a \textit{Kramers-Moyal} expansion of the (classical) master equation for the probability distribution of the process \cite{Risken:1984vl}. Below, we will restrict ourselves to certain forms of this generator, but for now only the Markov assumption will be used.

In order to derive an expression related to the reduced equations of
motion in a simple manner, consider the quantity $\hat{\rho}(t, \Omega) \equiv \hat{\rho}_\Omega(t)P(\Omega, t)$, which is the joint distribution of the quantum-mechanical coordinates and the stochastic variable at time $t$. $\hat{\rho}_\Omega(t) \equiv \hat{\rho}(t | \Omega(t))$ is the density matrix conditioned on the value of the stochastic process at time $t$ (in contrast to $\hat{\rho}(t|\{\Omega(t)\})$, which is the density matrix conditioned on an entire history of the stochastic process). Evaluating the time derivative of this joint distribution results in the \textit{stochastic Liouville equation}, first derived by Kubo \cite{Kubo:1969tj}:
\bqa
\frac{\partial \hat{\rho}(t, \Omega)}{\partial t} = \left[ -\frac{i}{\hbar}\hat{H}_0^\times -\frac{i}{\hbar}\Omega(t)\hat{V}^\times \right] \hat{\rho}(t, \Omega) + \Gamma_\Omega \hat{\rho}(t, \Omega),
\label{eq:stoch_l}
\eqa
where $A^\times B \equiv [A,B]$. Here, we have used the fact that because
the stochastic process is Markov, the time evolution of $P(\Omega, t)$ follows the same law as the conditional distribution in \erf{eq:markov_generator} and is generated by $\Gamma_\Omega$. 
This simple linear form for the evolution of the joint distribution is possible because both the conditional quantum density matrix and the probability distribution for the stochastic variable evolve linearly and in a Markov fashion. 
Because $\hat{\rho}(t,\Omega)$ is a joint distribution, its marginal over
the stochastic variable yields the average density matrix, i.e., $\hat{\rho}(t) = \int_\mathcal{D} \dd \Omega \hat{\rho}(t,\Omega)$, where $\mathcal{D}$ is the range of the stochastic variable \footnote{It is not immediately obvious that this integral gives the same average density matrix as the expression in \erf{eq:rho_functional} since it prescribes an integral over the variable at time $t$ whereas \erf{eq:rho_functional} prescribes a functional integral over the entire history of the stochastic process. This equivalence of the marginal derived from the solution of \erf{eq:stoch_l} and that given in \erf{eq:rho_functional} is entirely due to the Markovianity of the stochastic process and is proven in some detail in Ref. \cite{Ban:2010ja}.}.

{\it \textbf{Markov diffusion processes}}.
A wide a class of physical processes can be approximated by truncating
the Kramers-Moyal expansion for the generator $\Gamma_{\Omega}$ of the Markov process at the second term \cite{Risken:1984vl}. This results in the \textit{Fokker-Planck} equation for the probability distribution, with the generator being the differential operator \cite{Kam-2007}:
$\Gamma_\Omega^\textrm{FP} \equiv \Phi_\Omega = -
\frac{\partial}{\partial \Omega} A(\Omega) +
\frac{1}{2}\frac{\partial^2}{\partial\Omega^2}B(\Omega)$, 
where $A(\Omega)$ and $B(\Omega)$ are real differentiable coefficients,
with the restriction that $B>0$. 

For such \textit{diffusion processes}, it is common to define the backward, or adjoint, generator for the process:
$\overleftarrow{\Phi}_\Omega = A(\Omega) \frac{\partial}{\partial\Omega} + \frac{1}{2}B(\Omega)\frac{\partial^2}{\partial^2\Omega}$.
For a Markov process, \erf{eq:markov_generator} is often called the \textit{forward Kolmogorov equation} and $\frac{\partial}{\partial t'}P(\Omega, t | \Omega', t') = \overleftarrow{\Phi}_{\Omega'} P(\Omega, t | \Omega', t')$ is the \textit{backward Kolmogorov equation}. 
Let $f_n(\Omega)$ and $\overleftarrow{f}_n(\Omega)$ be eigenfunctions of
the forward and backward generators, respectively, i.e., $\Phi_\Omega
f_n(\Omega) = -\lambda_n f_n(\Omega)$ and $\overleftarrow{\Phi}_\Omega
\overleftarrow{f}_n(\Omega) = -\lambda_n \overleftarrow{f}_n(\Omega)$,
with $\lambda_n \geq 0$ \cite{Risken:1984vl}. These two eigenfunctions
are related through the stationary distribution of the process,
explicitly $f_n(\Omega) = P^0(\Omega) \overleftarrow{f}_n(\Omega)$,
where $P^{0}$ is the stationary distribution of the process $\Omega$
\cite{Gar-2004}. Furthermore, the eigenfunctions of the backward generator
form a complete, orthogonal system (on the range $\mathcal{D}$) under
the measure induced by the stationary distribution of the process:
$\int_\mathcal{D} \dd \Omega  P^0(\Omega) \overleftarrow{f}_n(\Omega)\overleftarrow{f}_m(\Omega) = \delta_{mn}$,
which in turn implies the following about the eigenfunctions of the forward generator \cite{Gar-2004}:
$\int_\mathcal{D} \dd \Omega  (P^0(\Omega))^{-1} f_n(\Omega) f_m(\Omega) =\delta_{mn}$

{\it \textbf{Diffusive hierarchical equations of motion}}.
Using the completeness and orthogonality of the backward and forward
generator eigenfunctions \cite{Gar-2004}, we can expand $P(\Omega, t)$ in terms of these eigenfunctions, and consequently,
$\hat{\rho}(t, \Omega) \equiv \hat{\rho}_\Omega(t)P(\Omega, t) = \sum_{n=0}^\infty \bar{\rho}_n(t) f_n(\Omega)$
Here, $\bar{\rho}_n(t)$ are unnormalized auxiliary density matrices,
which do not have the interpretation of being \emph{conditioned} density
matrices of the quantum system. Then, expressing the stochastic
Liouville equation \erf{eq:stoch_l} in terms of this eigenfunction
expansion yields
\bqa 
\sum_{n=0}^\infty \frac{\partial}{\partial t}\bar{\rho}_n(t) f_n(\Omega) = \sum_{n=0}^\infty -\frac{i}{\hbar}H_0^\times  \bar{\rho}_n(t) f_n(\Omega) \nn \\
- \frac{i}{\hbar}V^\times \bar{\rho}_n(t) \left[ \Omega(t) f_n(\Omega) \right] - \lambda_n \bar{\rho}_n(t) f_n(\Omega).
\eqa
To simplify further, we make the assumption that the eigenfunctions of
the backward Markov generator are polynomials in $\Omega$, i.e.,
$\overleftarrow{f}_n(\Omega)$ is an $n^\textrm{th}$ order polynomial in
$\Omega$. With this assumption, we utilize the following theorem,
typically attributed to Favard \cite{Schoutens:2000ux}, which states an
important property of orthogonal polynomials:

\textit{Theorem (Favard):} Let $\{p_n(x)\}, n\geq 0$ be a system of polynomials. This system satisfies a three-term recurrence relation $p_{n+1}(x) = (A_n x+ B_n)p_{n}(x) - C_n p_{n-1}(x)$,
(with $p_{-1}(x)=0)$ if and only if it is a system of \textit{orthogonal polynomials}. Here $A_n \neq 0, B_n$ and $C_n \neq 0$ are $\Omega$-independent (real) recurrence coefficients. 

Favard's theorem implies that the orthogonal polynomial eigenfunctions
of the backward Markov generator (and hence, those of the forward Markov
generator) satisfy the three-term recurrence relations:
\beq
\Omega f_n(\Omega) = \frac{C_n}{A_n}f_{n-1}(\Omega) - \frac{B_n}{A_n}f_n(\Omega) +\frac{1}{A_{n}}f_{n+1}(\Omega).
\eeq
Using this recurrence relation, the eigenfunction expansion of the stochastic Liouville equation becomes:
\begin{widetext}
\bqa 
\sum_{n=0}^\infty \frac{\partial}{\partial t}\bar{\rho}_n(t) f_n(\Omega) &=&  \sum_{n=0}^\infty \left[-\frac{i}{\hbar}H_0^\times - \lambda_n \right]   \bar{\rho}_n(t) f_n(\Omega)-\frac{i}{\hbar}V^\times \bar{\rho}_n(t) \left[  \frac{1}{A_{n}}f_{n+1}(\Omega) + \frac{C_n}{A_n}f_{n-1}(\Omega) - \frac{B_n}{A_n}f_n(\Omega) \right]
\eqa
Multiplying both sides of this equation by $f_m(\Omega)/P^0(\Omega)$ and integrating over $\Omega$ results in an equation for each $m$ because of the orthogonality of the functions $f_n(\Omega)$. These equations form a hierarchy of coupled (operator) differential equations, which we refer to as \textit{diffusive hierarchical equations of motion} (DHEOM) \footnote{The term \textit{diffusive} is included in this name to avoid confusion with the recently derived
  hierarchical equations of motion that describe average dynamics resulting from coupling to a fully quantum mechanical description of the environment \cite{Ishizaki:2005jw, *Ish.Fle-2009}.}:
\bqa
\frac{\partial}{\partial t}\bar{\rho}_0(t) &=& -\left[ \frac{i}{\hbar}\left( H_0^\times -\frac{B_0}{A_0} V^\times \right) + \lambda_0 \right]  \bar{\rho}_0(t)  -\frac{i}{\hbar} \frac{C_1}{A_1}\ V^\times \bar{\rho}_{1}(t), \nn \\
\frac{\partial}{\partial t}\bar{\rho}_n(t) &=& -\left[ \frac{i}{\hbar}\left( H_0^\times -\frac{B_n}{A_n} V^\times \right) + \lambda_n \right] \bar{\rho}_n(t) -\frac{i}{\hbar} \frac{C_{n+1}}{A_{n+1}}\ V^\times \bar{\rho}_{n+1}(t) - \frac{i}{\hbar}\frac{1}{A_{n-1}}  V^\times \bar{\rho}_{n-1}(t), ~~~~~~~ n > 0.
\label{eq:hierarchy}
\eqa
\end{widetext}
These equations are useful because they describe the evolution of the
system density matrix under the influence of stochastic noise, but
without explicit reference to noise variables. In addition, each member of the hierarchy only couples to its two neighbors, i.e., $n$ couples to $n+1$ and $n-1$, making their integration easy. We also need to specify
the initial conditions and the prescription for calculating the system
density matrix from the solution of the hierarchy of coupled
equations. A physically motivated initial state for the quantity
$\hat{\rho}(t, \Omega)$ is $\hat{\rho}(0, \Omega) = \hat{\rho}^0
P^0(\Omega)$, where $P^0$ is the stationary (equilibrium) distribution
of the noise process and $\hat{\rho}^0$ is any system density
matrix. Evolution from this initial state describes the response of a
quantum system to fluctuations around the bath equilibrium. Now,
$\Phi_\Omega P^0 = 0$ by definition, and therefore,
$\bar{\rho}_0(0) = \hat{\rho}^0$, and $\bar{\rho}_n(0)=0$ $\forall$ $n
\neq 0$. At any time, the system density matrix is defined as the
integral of the quantity $\hat{\rho}(t, \Omega)$ over $\Omega$:
\beq
\hat{\rho}(t) = \int_\mathcal{D} d\Omega ~ \hat{\rho}(t,\Omega) =  \int_\mathcal{D} d\Omega \sum_{n=0}^\infty \bar{\rho}_n(t) f_n(\Omega) \propto \bar{\rho}_0(t). \nn
\eeq
The proportionality is a result of the following property of the eigenfunctions:
\bqa
\int_\mathcal{D} d\Omega ~ f_n(\Omega) 
&\propto& \int_\mathcal{D}  d\Omega ~ P^0(\Omega) \overleftarrow{f}_n(\Omega) \overleftarrow{f}_0(\Omega) = \delta_{n0}, \nn
\eqa
where this proportionality follows from the fact that
$\overleftarrow{f}_0$ is a zeroth-order polynomial in $\Omega$. As we
shall see below, it is always possible to choose $\overleftarrow{f}_0 =
1$, and therefore, the proportionality above can be converted to an
equality. Hence, the auxiliary density matrix $\bar{\rho}_{0}(t)$ always
keeps track of the averaged system density matrix $\hat{\rho}(t)$.

To summarize, we have shown that the dynamics of a quantum system that
is linearly driven by a diffusion process with a Markov generator
possessing polynomial eigenfunctions can be described by a hierarchy of
coupled dynamical equations. The solution to these equations will
reproduce the reduced density matrix of the quantum system at any time,
with no approximations. However, the above hierarchy of equations is
infinite, and therefore, we require some truncation strategy in order to
solve them. If the equations can be truncated at some $n=N$ at the
expense of bounded error, they can be numerically solved (or
analytically solved via a partial fraction expansion if $H_0$ is
time-independent \cite{Tan-2006}). A general truncation strategy exists
if the following conditions hold: (1)  for large enough $n=N$,
$|\lambda_N| \gg ||H_0^\times - \frac{B_N}{A_N}V^\times||_2$, and (2)
$\frac{1}{A_n} \in \omega\left(\frac{C_{n}}{A_{n}}\right)$, where
$||\cdot ||_2$ is the induced 2-norm and $f_n \in \omega(g_n)$ is
indicates that $f_n$ dominates $g_n$ asymptotically (in $n$). In
appendix \ref{sec:truncation}, we develop such a general truncation strategy that results in the following \textit{terminator equation} for the hierarchy at a level $N$ satisfying these conditions:
\begin{widetext}
\beq
\frac{\partial}{\partial t}\bar{\rho}_N(t) = -\left[ \frac{i}{\hbar}\left( H_0^\times -\frac{B_N}{A_N} V^\times \right) + \lambda_N \right] \bar{\rho}_N(t) - \frac{1}{\hbar^2} \frac{C_{N+1}}{\lambda_{N+1} A_{N+1}A_N} V^\times V^\times\bar{\rho}_{N}(t) - \frac{i}{\hbar}\frac{1}{A_{N-1}}  V^\times \bar{\rho}_{N-1}(t).
\eeq
\end{widetext}

We now turn to characterizing the diffusive stochastic
processes with generators that have polynomial eigenfunctions. A quantum
system driven by each one of these processes will have a dynamical
description in terms of the DHEOM derived above. The combination of Bochner's
theorem \cite{Chihara:1978vz}, which is stated explicitly in appendix
\ref{sec:bochner}, and Favard's theorem implies that there are only
three diffusive processes with backward and forward generators that
have orthogonal polynomial eigenfunctions: (a) the Ornstein-Uhlenbeck
(OU) process, defined on $(-\infty, \infty)$ (b) the square-root
process, defined on $(\Omega_\textrm{min},\infty)$ and (c) the Jacobi
process, defined on $(\Omega_\textrm{min}, \Omega_\textrm{max})$. In
table \ref{tab:diffproc} of appendix \ref{sec:bochner}, we list these
diffusive processes with their orthogonal polynomial eigenfunctions and
other relevant properties. In appendix \ref{sec:heom_explicit}, we also explicitly write the DHEOM that describe the dynamics of a quantum system linearly driven by each of the above three stochastic processes. 
The Jacobi process is particularly efficient to simulate because the
eigenvalues $\lambda_n$ of its generator scale quadratically with $n$,
making truncation possible at a smaller depth than for the other two
processes, where $\lambda_n$ scales linearly with $n$. We note that the
DHEOM for the OU process was previously derived by Kubo and
Tanimura \cite{Kubo:1969tj, Tan-2006}, but this more general framework
resulting in DHEOMs for all major diffusive Markov processes was
heretofore unknown. For reasons explained in appendix
\ref{sec:heom_explicit}, the DHEOM method is limited to simulating
square-root processes with mean reversion rate $\gamma>1$, whereas any
$\gamma$ is valid for the other processes. Finally, for completeness
appendix \ref{sec:hierarchy_map} presents the DHEOM for the propagator
for the dynamical map as opposed to the density matrix.

{\it \textbf{Application and Discussion}}.
\begin{figure}[t]
\centering
\includegraphics[width=8.5cm]{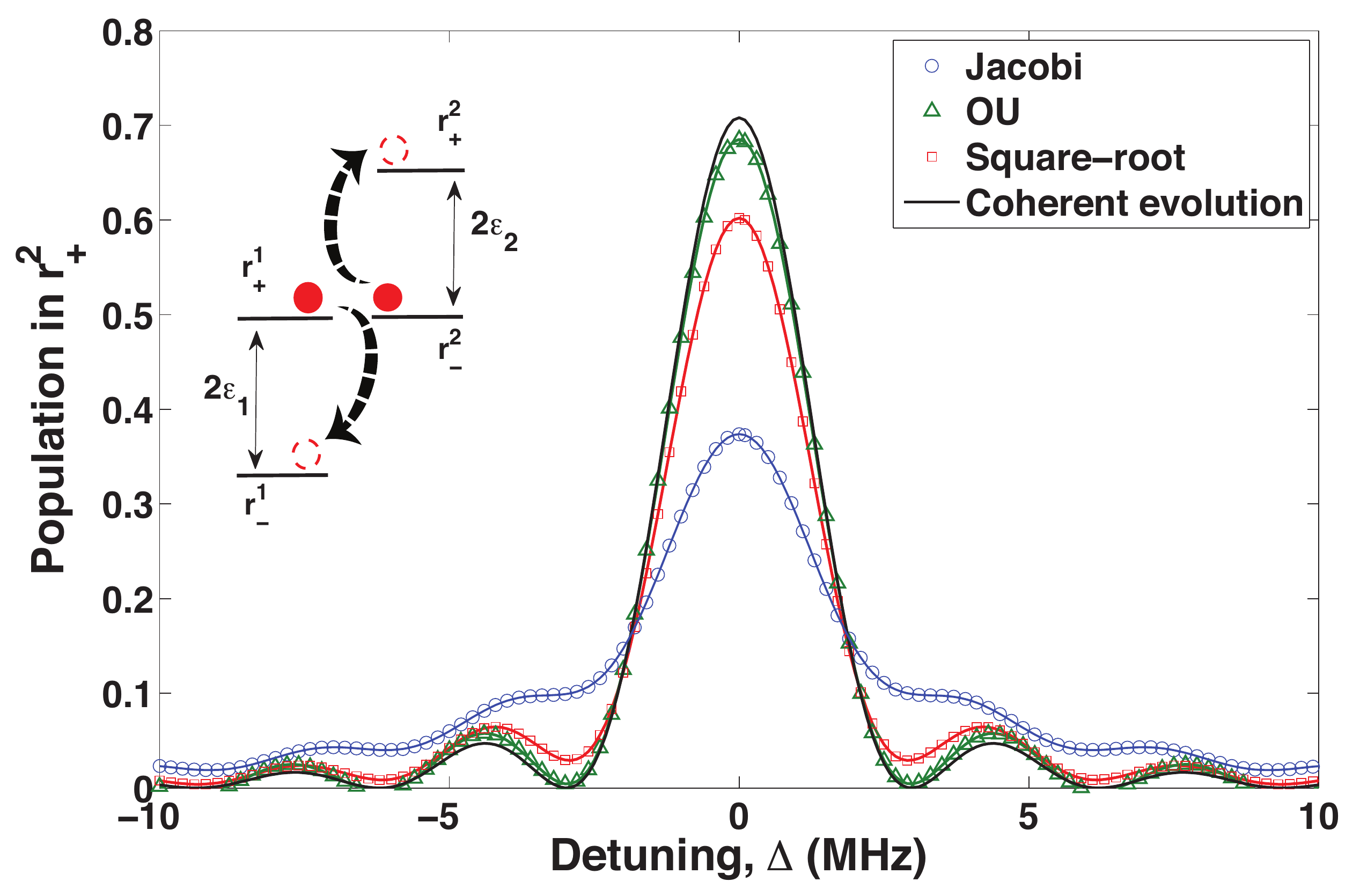}
\caption{Average population in the excited state of atom 2 after an
  interaction time of $T=1\mu$s as a function of the Stark detuning for
  different models of motional noise. The black (solid) curve is for
  coherent evolution with no fluctuation in inter-atom distance. 
  The parameters used are: $\gamma=1.5, \mu=1$ for all noise sources, and $\sigma^2=0.3$ for the
  OU process. The square root process is defined in the semi-interval
  $[0,\infty)$ with $c_1=1$, and the Jacobi process in the interval
    $(1/8,8)$ with $c=1$. See appendix \ref{sec:bochner} for the interpretation of these parameters.
\label{fig:rydberg_eg} }
\end{figure}
The primary advantage of our general formulation of DHEOM is that
it allows for the efficient simulation of quantum systems driven by
noise sources with bounded range and non-Gaussian amplitude
distribution. As an application, we consider energy
transfer between two systems mediated by a dipole-dipole
interaction. This interaction is an approximation to the
Coulomb coupling of electric charge distributions and is the basis of
several common experimental techniques, such as F\"orster resonance
energy transfer (FRET) spectroscopy \cite{Sahoo:2011jt} and Stark-tuned resonant
transfer in Rydberg atoms \cite{vanDitzhuijzen:2008ch, Ryabtsev:2010do};
we focus on this latter system as our example.

The electric dipole-dipole coupling energy scales as $J \propto 1/R^3$,
where $R$ is the distance between the coupled molecules or
atoms. Consider the case where this inter-particle spacing is varying in
time as a result of thermal, electromagnetic, or vibrational
fluctuations. If the fluctuations in $R$ are treated perturbatively the
dipole-dipole coupling energy can be expanded linearly around the mean
separation, $R_0$: $J \sim 1/R_0^3 - (3/R_0^2) \delta R(t)$. However, in
cases where such a perturbative treatment is not accurate
we must consider the energy $J$ as a time-dependent fluctuating quantity
in a range $[J_\textrm{min}, J_\textrm{max}]$. To explore the
consequences of such a non-perturbative treatment, consider the
Hamiltonian of two interacting Rydberg atoms:
$H = -\epsilon_1 \sigma_z^1 -\epsilon_2 \sigma_z^2 + J(t)(\sigma_+^1 \sigma_-^2 + \sigma_-^1 \sigma_+^2)$,
where we restrict the description of the atoms to the two energy levels
relevant to the energy transfer, and therefore use Pauli matrices to
describe them as effective two-level systems in the basis $\ket{r_-^i},
\ket{r_+^i}$ where $r_\pm^i$ are Rydberg states for atom $i$ with the
energy gap $\epsilon_i$, i.e., $\sigma_{z}^{i} \ket{r_{\pm}^i} = \pm
\epsilon_{i} \ket{r_{\pm}^i}$. Typically, these levels are chosen so
that when an electric field tunes atomic energies by the Stark effect they satisfy $\epsilon_1 = \epsilon_2$ (F\"orster resonance
condition) at some critical field value. For example, in
Ref.~\cite{Ryabtsev:2010do},
$\ket{r^1_+} = \ket{37P_{3/2}}, \ket{r^1_-} = \ket{37S_{1/2}},
\ket{r^2_+} = \ket{38S_{1/2}}, \ket{r^2_-} = \ket{37P_{3/2}}$, where
these are fine states of Rb atoms. We ignore relaxation of these states because for sufficiently cooled Rydberg atoms it can be neglected on the timescales we are considering \cite{Saf.Wal.etal-2010}. We also assume
that the relative orientation of the atomic dipoles remains
constant. The initial state is the excited state of atom 1: $\rho^0 =
\ket{r_+^1, r_-^2}\bra{r_+^1, r_-^2}$, and the dipole-dipole coupling is assumed to
fluctuate as $J(t) \equiv J_0\Omega(t)$, where $\Omega(t)$ is one of the
three diffusive noise processes described above. The quantity being observed (e.g., by selective field ionization \cite{Ryabtsev:2010do}) is the population in the excited state of atom 2 after an interaction time
$T$. Although position fluctuations were reported to be minor in the experiment of Ref. \cite{Ryabtsev:2010do}, we use physical parameters compatible with this experiment for concreteness: $J_0/\hbar = 0.5$MHz, and $T=1 \mu$s. Fig.~\ref{fig:rydberg_eg} shows the average population of
$\ket{r_+^2}$ at $T$ for different values of the Stark detuning $\Delta \equiv
\epsilon_2-\epsilon_1$. We expect that this transferred population will drop off as $|\Delta |$ increases (as the atoms become less resonant), but the figure shows that the behavior of this drop-off depends heavily on the exact nature of the fluctuations in $R$, although all
processes have the same mean $\mu=1$ and mean reversion rate $\gamma=1.5$. The OU and square-root processes
predict average populations that have significant oscillations as a
function of the detuning, similar to the completely coherent (noiseless) case. In
contrast, the Jacobi process driven evolution predicts damped oscillations and smaller transfer populations. We predict that for experiments performed in this regime, where position fluctuations are relevant, the Jacobi process average will be more accurate (and they are certainly more consistent with what is experimentally observed in Ref. \cite{Ryabtsev:2010do}). This is because the Jacobi process is a more accurate description than the OU and square-root processes, which can only be approximations in this scenario; realistic trapping conditions impose strict lower and upper
bounds on $J$ resulting from the fact that the atoms can only drift
within the trap volumes. The OU and square-root simulations predict
large transfer populations because rare, unphysical, large magnitude
couplings contribute to the averages in these cases while they do not to
the Jacobi process average. These results emphasize the advantage of
being able to efficiently average over non-Gaussian noise processes with
bounded range. To underscore the numerical efficiency of the DHEOM
simulations, we note that the calculations producing
Fig.~\ref{fig:rydberg_eg} were performed 10 times faster using the
present theory than conventional Monte Carlo sampling of the noise
processes \footnote{The DHEOM simulations required 1984 seconds on a
  standard laptop, while a Monte Carlo simulation (with 500 samplings
  over the noise processes to obtain reasonable convergence of the
  average) required 19403 seconds on the same machine. Both codes were
  unoptimized.}. Furthermore, because of the well-defined truncation
strategy, convergence issues and noise are not a concern for DHEOM
simulations as in the Monte Carlo approach.

{\it \textbf{Summary}}.
We have completely characterized the class of diffusively driven quantum systems for which a useful hierarchy of reduced equations of motion can be derived, and explicitly derived these DHEOM. We applied the technique to examine position fluctuation dependence of Stark tuned F\"orster resonance between Rydberg atoms. We expect that the general techniques developed in this work will be useful for widening the range of open quantum systems that can be simulated efficiently. Use of the DHEOM for Jacobi processes will be particularly useful for modeling quantum systems subject to fluctuations with bounded range, e.g., classical noise in external control fields \cite{Grace07a, *Grace07b, *Gra.Dom.etal-2010, *Young10a}. 

{\it \textbf{Acknowledgements}}. 
We acknowledge helpful discussions with Constantin Brif and Kevin Young
(SNL-CA). This work was supported by the Laboratory Directed Research
and Development program at Sandia National Laboratories. Sandia is a
multi-program laboratory managed and operated by Sandia Corporation, a
wholly owned subsidiary of Lockheed Martin Corporation, for the United
States Department of Energy's National Nuclear Security Administration
under contract DE-AC04-94AL85000. 

\bibliography{./hierarchy}
\bibliographystyle{apsrev4-1}

\appendix

\begin{widetext}
\pagebreak
\section{Bochner's theorem and polynomial eigenfunctions}
\label{sec:bochner}
In the main text, we claim that there are three stochastic processes that admit an eigenfunction of their Markov generator that is an orthogonal polynomial. This result follows from Bochner's theorem \cite{Chihara:1978vz}:

\textit{Theorem (Bochner):} Consider the eigenvalue problem
\beq
f(x)\frac{d^2}{dx^2}y(x) + g(x) \frac{d}{dx}y(x) + h(x) y(x) = \lambda y(x)
\label{eq:bochner}
\eeq
where $f, g, h$ are fixed polynomials and $\lambda$ is independent of $x$. There are only five classes of polynomial solutions to this problem, where $y_n(x)$ is a polynomial of degree $n$, and they are:
\begin{enumerate}
\item $y_n(x)=\textrm{He}_n(x), n\in\mathbb{N}_0$ is a Hermite polynomial, defined on the infinite interval $x\in (-\infty, \infty)$.
\item $y_n(x)=L^{(\alpha)}_n(x), n\in \mathbb{N}_0$ is a generalized Laguerre polynomial, defined on the semi-infinite interval $x\in [a, \infty)$ or $x\in (-\infty, a]$ for $a \in \mathbb{R}$. 
\item $y_n(x)=J^{(\alpha, \beta)}_n(x), n\in \mathbb{N}_0$ is a Jacobi polynomial, defined on the interval $x\in (a, b)$ for $a,b \in \mathbb{R}$.
\item $y_n(x)$ is in the set $\{x^n\}$ for $n\in \mathbb{N}_0$.
\item $y_n(x)=B_n(x)$ is a Bessel polynomial, defined on the infinite interval $x\in (-\infty, \infty)$.
\end{enumerate}
Note that the backward Kolmogorov equation is an example of the
differential equation considered in \erf{eq:bochner}, with the
additional restriction that $f(x)> 0$. Therefore, all polynomial
eigenfunctions of the backwards Kolmogorov generator fall within the
Bochner classes identified above. However, since we also require that
our eigenfunctions satisfy a three-term recurrence relation Favard's
theorem further restricts us to the subset of the Bochner classes that
are orthogonal polynomials. The Hermite, Laguerre, and Jacobi
polynomials are all classical orthogonal polynomials and therefore
obviously orthogonal. The polynomials sequence in class 4 above is not
an orthogonal sequence, and similarly, the Bessel polynomials are not an
orthogonal set of polynomials for any positive-definite measure. The
Bessel polynomials are orthogonal under a quasi-definite measure and
hence do satisfy a three-term recurrence relation
\cite{Chihara:1978vz}. However the positive-definiteness  of the measure
is an important feature in the following and therefore we will not
consider the Bessel polynomials. Therefore, the combination Favard's and
Bochner's theorems imply that there are only three diffusive processes
with backward and forward generators that have orthogonal polynomial eigenfunctions: (a) the Ornstein-Uhlenbeck process, (b) the square-root (or Cox-Ingersoll-Ross) process, and (c) the Jacobi process. This result has been derived by a number of authors in the past, in different mathematical contexts, including in Refs. \cite{Wong:1962tg, *Mazet:1997wi, *Gourieroux:2007vq}. 

\begin{center}
\begin{table}[]
\caption{\label{tab:diffproc}Summary of diffusion processes with
  backward generators possessing polynomial eigenfunctions. $I_{(a,b)}$
  is the indicator function for the interval $(a,b)$, $\Gamma[\cdot]$ is
  the Gamma function, and $B[\cdot]$ is the Beta function. All processes have exponentially decaying correlation functions, proportional to $e^{-\gamma t}$. Note that in order to get convenient recurrence relations, we choose unconventional
  normalizations for the orthogonal polynomial eigenfunctions.}
\begin{tabular}{|l|}
\hline
\textbf{Ornstein-Uhlenbeck process} \\
\hline
\textsf{Range of stochastic process} \\
\spa $\mathcal{D}_{\textrm{OU}}: -\infty < \Omega < \infty$ \\
\textsf{Generator parameters}  \\
\spa $A^\textrm{OU}(\Omega) = -\gamma(\Omega-\mu), ~~~ \gamma>0, \mu \in \mathcal{D}_{\textrm{OU}}$ \\
\spa $B^\textrm{OU}(\Omega) = \sigma^2$ \\
\textsf{Eigenfunctions of backward generator} \\
\spa Scaled Hermite polynomials \\
\spa $\overleftarrow{\phi}_n(\Omega) \equiv \frac{1}{n!}\textrm{He}_n(\sqrt{\frac{2\gamma}{\sigma^2}}(\Omega-\mu) ) = \sum_{m=0}^{\lfloor n/2 \rfloor} \frac{(-1)^m}{m! 2^m (n-2m)!} \left( \sqrt{\frac{2\gamma}{\sigma^2}}(\Omega-\mu)\right)^{n-2m}$ \\
\textsf{Eigenvalues of backward generator} \\
\spa $\overleftarrow{\Phi}^{\textrm{OU}}_\Omega \overleftarrow{\phi}_n(\Omega) = -n\gamma \overleftarrow{\phi}_n(\Omega)$; ~~~~~ $\lambda_n = n\gamma$ \\
\textsf{Stationary distribution} \\
\spa Gaussian distribution
\spa $P^\textrm{OU}_0(\Omega) =  \frac{1}{\sqrt{\pi \sigma^2/\gamma}} \textrm{e}^{-\frac{\gamma(\Omega-\mu)^2}{\sigma^2}}$ \\
\hline
\hline
\textbf{Square-root (or Cox-Ingersoll-Ross) process}\\
\hline
\textsf{Range of stochastic process} \\
\spa $\mathcal{D}_\textrm{SR}: -\frac{c_0}{c_1} < \Omega < \infty, ~~~ c_1>0$ \\
\textsf{Generator parameters}  \\
\spa $A^\textrm{SR}(\Omega) = -\gamma(\Omega-\mu), ~~~ \gamma>0, \mu \in \mathcal{D}_\textrm{SR}$ \\
\spa $B^\textrm{SR}(\Omega) = c_1 \Omega + c_0$ \\
\textsf{Eigenfunctions of backward generator} \\
\spa Scaled generalized Laguerre polynomials \\
\spa $\overleftarrow{\chi}_n(\Omega) \equiv \frac{1}{n!} L^{(\alpha)}_n \left[ \frac{2\gamma}{c_1}(\Omega+\frac{c_0}{c_1}) \right] =  \frac{1}{n!} \sum_{m=0}^n {n+\alpha \choose n-m} \frac{ \left( \frac{-2\gamma}{c_1}(\Omega+\frac{c_0}{c_1}) \right)^m}{m!} $ \\
\spa \spa where $\alpha \equiv \frac{2\gamma}{c_1}\left(\mu + \frac{c_0}{c_1} \right)-1$ \\
\textsf{Eigenvalues of backward generator} \\
\spa $\overleftarrow{\Phi}^{\textrm{SR}}_\Omega \overleftarrow{\chi}_n(\Omega) = -n\gamma \overleftarrow{\chi}_n(\Omega)$; ~~~~~ $\lambda_n = n\gamma$ \\
\textsf{Stationary distribution} \\
\spa  Gamma distribution
\spa $P_0^\textrm{SR}(\Omega) = \frac{\left(\frac{2\gamma}{c_1}\right)^{\alpha+1}}{\Gamma\left[\alpha+1\right]} \textrm{e}^{-\frac{2\gamma}{c_1}\left(\Omega+\frac{c_0}{c_1}\right)} \left( \Omega + \frac{c_0}{c_1}\right)^\alpha I_{(\frac{-c_0}{c_1}, \infty)}$  \\
\hline
\hline
\textbf{Jacobi process}\\
\hline
\textsf{Range of stochastic process} \\
\spa $\mathcal{D}_\textrm{J}: \omega_1 < \Omega < \omega_2$ \\
\textsf{Generator parameters}  \\
\spa $A^\textrm{J}(\Omega) = -\gamma(\Omega-\mu), ~~~ \gamma>0, \mu \in \mathcal{D}_\textrm{J}$ \\
\spa $B^\textrm{J}(\Omega) = -c(\Omega-\omega_1)(\Omega-\omega_2), ~~~ c>0$ \\
\textsf{Eigenfunctions of backward generator} \\
\spa Scaled Jacobi polynomials \\
\spa $\overleftarrow{\zeta}_n(\Omega) \equiv \frac{1}{n!} J^{(\alpha,\beta)}_n(\frac{2(\Omega-\omega_2)}{\omega_2-\omega_1}+1) $ \\ 
\spa\spa~~~ $= \frac{1}{n!} \sum_{m=0}^n {n \choose m} \frac{\Gamma[\alpha+\beta+n+m+1]}{\Gamma[\alpha+m+1]}\left(\frac{\Omega-\omega_2}{\omega_2-\omega_1}\right)^m$ \\
\spa \spa where $\alpha \equiv \frac{2\gamma}{c}\frac{\omega_2-\mu}{\omega_2-\omega_1}-1$, and $\beta \equiv \frac{2\gamma}{c}\frac{\mu-\omega_1}{\omega_2-\omega_1}-1$ \\
\textsf{Eigenvalues of backward generator} \\
\spa $\overleftarrow{\Phi}^{\textrm{J}}_\Omega \overleftarrow{\zeta}_n(\Omega) = (-n\gamma-\frac{1}{2}cn(n-1)) \overleftarrow{\zeta}_n(\Omega)$; ~~~~~ $\lambda_n = n\gamma+\frac{1}{2}cn(n-1)$ \\
\textsf{Stationary distribution} \\
\spa Beta distribution
\spa $P^{\textrm{J}}_0(\Omega) = \frac{1}{B[\beta+1,\alpha+1]}\frac{(\Omega-\omega_1)^\beta (\omega_2-\Omega)^\alpha}{(\omega_2-\omega_1)^{2\gamma /c-1}} I_{(\omega_1,\omega_2)}$ \\
\hline
\end{tabular}
\label{default}
\end{table}
\end{center}

\section{Truncation of DHEOM}
\label{sec:truncation}

The conditions for a general truncation strategy to exist for the DHEOM are: (1)  for large enough $n=N$, $|\lambda_N| \gg ||H_0^\times - \frac{B_N}{A_N}V^\times||_2$, and (2) $\frac{1}{A_n} \in \omega\left(\frac{C_{n}}{A_{n}}\right)$, where $||\cdot ||_2$ is the operator 2-norm and $f_n \in \omega(g_n)$ is indicates that $f_n$ dominates $g_n$ asymptotically (in $n$). To develop the truncation strategy under these conditions we follow Ref. \cite{Tan-2006} and consider the formal solution to the $n=N+1 \gg 1$ member of the hierarchy:
\bqa
\bar{\rho}_{N+1}(t) = \frac{-i}{\hbar}\int_0^t \dd{\tau} \exp\left\{ -\left[\frac{i}{\hbar}\left(H_0^\times - \frac{B_{N+1}}{A_{N+1}}V^\times \right) + \lambda_{N+1} \right](t-\tau)\right\} V^\times \left[ \frac{C_{N+2}}{A_{N+2}}\bar{\rho}_{N+2}(\tau) + \frac{1}{A_{N}} \bar{\rho}_{N}(\tau)\right] 
\eqa
The first condition above implies that by choice of $N$ we have $|\lambda_{N+1}| \gg ||H_0 - \frac{B_{N+1}}{A_{N+1}}V||$, and we can replace $\bar{\rho}_i(\tau)$ with $\bar{\rho}_i(t)$ in the integrand because the exponential term decays much faster than the rate at which $\bar{\rho}_i(\tau)$ changes. This approximation enables the integral to be computed and results in the solution
\beq
\bar{\rho}_{N+1}(t) \approx \frac{-i}{\hbar \lambda_{N+1}}V^\times \left[ \frac{C_{N+2}}{A_{N+2}}\bar{\rho}_{N+2}(t) + \frac{1}{A_{N}}\bar{\rho}_{N}(t)\right] \nn
\eeq
Now the second condition above implies that for large enough $N$, $\frac{C_{N+2}}{A_{N+2}} \ll  \frac{1}{A_{N}}$ and we can ignore the contribution from the first term in the square bracket, resulting in:
$\bar{\rho}_{N+1}(t) \approx  \frac{-i}{\hbar \lambda_{N+1} A_{N}}V^\times\bar{\rho}_{N}(t)
$. Using this approximation, we can terminate the hierarchy at level $N$ and formulate the \textit{terminator} equation as:
\beq
\frac{\partial}{\partial t}\bar{\rho}_N(t) = -\left[ \frac{i}{\hbar}\left( H_0^\times -\frac{B_N}{A_N} V^\times \right) + \lambda_N \right] \bar{\rho}_N(t) - \frac{1}{\hbar^2} \frac{C_{N+1}}{\lambda_{N+1} A_{N+1}A_N} V^\times V^\times\bar{\rho}_{N}(t) - \frac{i}{\hbar}\frac{1}{A_{N-1}}  V^\times \bar{\rho}_{N-1}(t)
\eeq

\section{Explicit forms of the DHEOMs}
\label{sec:heom_explicit}

In this Appendix we explicitly write DHEOM for quantum systems driven by the three diffusive Markov processes that permit such descriptions. We have confirmed that these DHEOM descriptions of dynamical evolution agree with averaged Monte-Carlo simulations of the same noise processes. 

\subsection{Ornstein-Uhlenbeck process}
For the case where $\Omega(t)$ is an Ornstein-Uhlenbeck process, Kubo and Tanimura \cite{Kubo:1969tj, Tan-2006} have demonstrated how to construct hierarchical equations of motion for the reduced density matrix. In fact, the generalization described in this paper was originally motivated by this work. In this section we restate the Kubo-Tanimura DHEOM for the OU process in the present notation. 

The joint distribution is first expanded in terms of the eigenfunctions of the Markov generator for the Ornstein-Uhlenbeck process:
\beq
\hat{\rho}(t, \Omega) \equiv \hat{\rho}_\Omega(t)P(\Omega, t) = \sum_{n=0}^\infty \bar{\rho}_n(t) \phi_n(\Omega)
\eeq
 $\phi_n(\Omega) = P^{\textrm{OU}}_0(\Omega)\overleftarrow{\phi}_n(\Omega)$ are eigenfunctions of the forward generator for the OU process. These quantities are defined explicitly in Table \ref{tab:diffproc}. Since $\phi_n(\Omega)$ are proportional to Hermite polynomials a three term recurrence relation they follow is:
\beq
\Omega \phi_n(\Omega) = \sqrt{\frac{\sigma^2}{2\gamma}} \phi_{n-1}(\Omega) + \mu \phi_n(\Omega) + (n+1)\sqrt{\frac{\sigma^2}{2\gamma}}\phi_{n+1}(\Omega) 
\label{eq:hermite_recur}
\eeq
where Table \ref{tab:diffproc} should be referred to for the interpretation of the parameters in this expression. 
Given this recurrence relation, the resulting DHEOM is:
\bqa
\frac{\partial}{\partial t}\bar{\rho}_0(t) &=& -\frac{i}{\hbar}\left(H_0^\times +\mu V^\times \right)  \bar{\rho}_0(t)  -\frac{i}{\hbar} \sqrt{\frac{\sigma^2}{2\gamma}}\ V^\times \bar{\rho}_{1}(t), \nn \\
\frac{\partial}{\partial t}\bar{\rho}_n(t) &=& -\left[ \frac{i}{\hbar}\left( H_0^\times +\mu V^\times \right) + n\gamma \right] \bar{\rho}_n(t) -\frac{i}{\hbar} \sqrt{\frac{\sigma^2}{2\gamma}}\ V^\times \bar{\rho}_{n+1}(t) - \frac{i}{\hbar}n \sqrt{\frac{\sigma^2}{2\gamma}}\ V^\times \bar{\rho}_{n-1}(t), ~~~~~~~ n > 0
\label{eq:hierarchy}
\eqa
The terminator equation is explicitly, 
\beq
\frac{\partial}{\partial t}\bar{\rho}_N(t) = -\left[\frac{i}{\hbar} \left( H_0^\times + \mu V^\times \right)+ N\gamma \right] \bar{\rho}_N(t) - \frac{1}{\hbar^2\gamma} \left(\frac{\sigma^2}{2\gamma}\right) V^\times V^\times \bar{\rho}_{N}(t) - \frac{i}{\hbar}N \sqrt{\frac{\sigma^2}{2\gamma}}V^\times \bar{\rho}_{N-1}(t)
\eeq

\subsection{Square-root process}
We will expand the joint distribution for the quantum and stochastic variables as
\beq
\hat{\rho}(t, \Omega) = \sum_{n=0}^\infty \bar{\rho}_n(t) \chi_n(\Omega)
\eeq
where $\chi_n(\Omega) = P^{\textrm{SR}}_0(\Omega)\overleftarrow{\chi}_n(\Omega)$ is the expression for the eigenfunctions of the forward generator for the square-root process in terms of the quantities in Table \ref{tab:diffproc}. The three-term recurrence formula satisfied by the basis functions $\chi_n(\Omega)$ (which follows from a three-term recurrence relation followed by the Laguerre polynomials) that we require is:
\bqa
\Omega \chi_n(\Omega) &=& -\frac{c_1}{2\gamma}\frac{(\alpha+n)}{n} \chi_{n-1}(\Omega) \nn \\
&& + ~\left[ \frac{c_1}{2\gamma}(\alpha+2n+1)-\frac{c_0}{c_1} \right]\chi_n(\Omega) \nn \\
&& - ~ \frac{c_1}{2\gamma}(n+1)^2 \chi_{n+1}(\Omega)
\label{eq:laguerre_recur}
\eqa
where $\alpha \equiv \frac{2\gamma}{c_1}\left(\mu + \frac{c_0}{c_1} \right)-1$, and $\gamma, \mu, c_0,$ and $c_1$ are the defining parameters for the process (see Table \ref{tab:diffproc}). Using this recurrence relation, the resulting hierarchy of coupled equations of motion are:
\bqa
\frac{\partial}{\partial t}\bar{\rho}_0(t) &=& -\frac{i}{\hbar}\left(H_0^\times + \left[\frac{c_1}{2\gamma}(\alpha+1)-\frac{c_0}{c_1}\right] V^\times \right)  \bar{\rho}_0(t)  +\frac{i}{\hbar} \frac{c_1}{2\gamma}(\alpha+1) V^\times \bar{\rho}_{1}(t), \nn \\
\frac{\partial}{\partial t}\bar{\rho}_n(t) &=& -\left[\frac{i}{\hbar}\left( H_0^\times + \left[ \frac{c_1}{2\gamma}(\alpha+2n+1)-\frac{c_0}{c_1} \right] V^\times\right) + n\gamma \right] \bar{\rho}_n(t)  \nn \\
&+& \frac{i}{\hbar}\frac{c_1}{2\gamma}\frac{(\alpha+n+1)}{n+1} V^\times \bar{\rho}_{n+1}(t) + \frac{i}{\hbar} \frac{c_1}{2\gamma}n^2 V^\times \bar{\rho}_{n-1}(t), ~~~~~~~ n > 0
\label{eq:hierarchy_sr}
\eqa
The terminator equation is explicitly,
\bqa
\frac{\partial}{\partial t}\bar{\rho}_N(t) &=& -\left[\frac{i}{\hbar}\left( H_0^\times + \left[ \frac{c_1}{2\gamma}(\alpha+2N+1)-\frac{c_0}{c_1} \right] V^\times\right) + N\gamma \right] \bar{\rho}_N(t)  \nn \\
&-& \frac{1}{\hbar^2 \gamma} \left(\frac{c_1}{2\gamma} \right)^2 (\alpha+N+1) V^\times V^\times \bar{\rho}_{N}(t) + \frac{i}{\hbar} \frac{c_1}{2\gamma} N^2 V^\times \bar{\rho}_{N-1}(t)
\eqa

Note that this choice of recurrence relation leads to the coefficient of $V^\times$ scaling as $\rightarrow n/\gamma$ for large $n$. At the same time, the eigenvalue of the generator scales as $\rightarrow \gamma n$. This means that the first condition for the existence of a truncation strategy, that $|\lambda_N| \gg ||H_0^\times - \frac{B_N}{A_N}V^\times||_2$ can only hold true for $\gamma>1$. Therefore we are only guaranteed that a valid truncation exists when $\gamma>1$ for the square-root process. It is possible that this condition can be lifted by using an alternate recurrence relation for the generalized Laguerre polynomials but we have not been able to find such a recurrence relation at this time, and therefore are limited to simulating square-root processes with $\gamma>1$ using the above DHEOM.

\subsection{Jacobi process}
If the stochastic driving of the quantum system is by a Jacobi process, we expand the joint distribution in terms of the eigenfunctions of its generator
\beq
\hat{\rho}(t, \Omega) = \sum_{n=0}^\infty \bar{\rho}_n(t) \zeta_n(\Omega)
\eeq
where $\zeta_n(\Omega) = P^\textrm{J}_0(\Omega)\overleftarrow{\zeta}_n(\Omega)$ in terms of the quantities defined in Table \ref{tab:diffproc}. The three-term recurrence formula we will use, which follows from a three-term recurrence formula of Jacobi polynomials, is:
\bqa
\Omega \zeta_n(\Omega) &=& \frac{\Delta\omega}{2}\left(\frac{(\alpha+n)(\beta+n)}{n \eta_n(\eta_n+1)}\right) \zeta_{n-1}(\Omega) \nn \\
&+&\left[\omega_2 - \frac{\Delta\omega}{2}\left(\frac{\alpha^2-\beta^2}{\eta_n(\eta_n+2)}+1\right) \right] \zeta_n(\Omega) \nn \\
&+& \frac{\Delta\omega}{2}\left(\frac{(n+1)^2(\eta_n-n+1)}{(\eta_n+1)(\eta_n+2)}\right) \zeta_{n+1}(\Omega) 
\eqa
where $\alpha$ and $\beta$ are as defined in Table \ref{tab:diffproc}, and $\Delta\omega\equiv \omega_2-\omega_1$, and $\eta_n \equiv \alpha+\beta+2n$. Using this recurrence relation, we derive the following DHEOM: 
\bqa
\frac{\partial}{\partial t}\bar{\rho}_0(t) &=& -\frac{i}{\hbar}\left(H_0^\times + \left[\omega_2 - \frac{\Delta\omega}{2}\left(\frac{\alpha^2-\beta^2}{\eta_0(\eta_0+2)}+1\right)\right]V^\times \right) \bar{\rho}_0(t) - \frac{i}{\hbar} \frac{\Delta\omega}{2}\left(\frac{(\alpha+1)(\beta+1)}{\eta_1(\eta_1+1)}\right) V^\times \bar{\rho}_1(t)  \nn \\
\frac{\partial}{\partial t}\bar{\rho}_n(t) &=& -\left[\frac{i}{\hbar}\left(H_0^\times + \left[\omega_2-\frac{\Delta\omega}{2}\left(\frac{\alpha^2-\beta^2}{\eta_n(\eta_n+2)}+1\right)\right]V^\times\right)+(n\gamma+\frac{1}{2}cn(n-1)) \right] \bar{\rho}_n(t) \nn \\
&& ~~~~ - \frac{i}{\hbar} \Delta\omega\left(\frac{(\alpha+n+1)(\beta+n+1)}{(n+1)\eta_{n+1}(\eta_{n+1}+1)}\right) V^\times \bar{\rho}_{n+1}(t)  \nn \\
&& ~~~~ - \frac{i}{\hbar} \Delta\omega \left(\frac{n^2(\eta_{n-1}-n+2)}{(\eta_{n-1}+1)(\eta_{n-1}+2)}\right) V^\times \bar{\rho}_{n-1}(t), ~~~~~~~~~~~ n > 0
\label{eq:hierarchy_sr}
\eqa
The terminator equation for this DHEOM is explicitly,
\bqa
\frac{\partial}{\partial t}\bar{\rho}_N(t) &=& -\left[\frac{i}{\hbar}\left(H_0^\times + \left[\omega_2-\frac{\Delta\omega}{2}\left(\frac{\alpha^2-\beta^2}{\eta_N(\eta_N+2)}+1\right)\right]V^\times\right)+(N\gamma+\frac{1}{2}cN(N-1)) \right] \bar{\rho}_N(t) \nn \\
&& ~~~~ - \frac{1}{\hbar^2(\gamma + \frac{c}{2}N)} \Delta \omega^2 \left(\frac{(\alpha+N+1)(\beta+N+1)(\eta_N-N+1)}{\eta_{N+1}(\eta_{N}+1)(\eta_{N+1}+1)(\eta_N+2)}\right) V^\times V^\times \bar{\rho}_{N}(t)  \nn \\
&& ~~~~ - \frac{i}{\hbar} \Delta\omega \left(\frac{N^2(\eta_{N-1}-N+2)}{(\eta_{N-1}+1)(\eta_{N-1}+2)}\right) V^\times \bar{\rho}_{N-1}(t)
\eqa

It is interesting to note that the growth of the $\lambda_n$ with $n$ for the Jacobi process is quadratic (as opposed to linear for the OU and square-root processes). This implies that the termination conditions for the DHEOM are met much more quickly for the Jacobi process, and hence it is more efficient to simulate than the other processes.  

\section{DHEOM for the integrated map}
\label{sec:hierarchy_map}
In the main text, we derived a general DHEOM for the density matrix
describing a quantum system driven by a diffusive Markov process. In
solving for the dynamics of a quantum systems, it is often useful to
work at the level of an integrated (completely positive) map, or
propagator, rather than a density matrix. The two are related simply as
$\rho(t) = \mathcal{E}(t)\rho^0$, where $\mathcal{E}(t)$ is the
integrated map that propagates any initial state $\rho^0$ to
$\rho(t)$. If one chooses to vectorize the $n\times n$ density matrix
into an $n^{2} \times 1$ vector by stacking the columns:
$\overrightarrow{\rho} = \textrm{vec}({\hat{\rho}})$, then
$\mathcal{E}(t)$ is a matrix of size $n^2 \times n^2$. For unitary
evolution, where $\rho(t) = U(t)\rho_0 U\dg(t)$, we get a matrix
representation $\mathcal{E}(t) = U^*(t)\otimes U(t)$ from the identity
$\textrm{vec}(ABC) = (C^T \otimes A) \textrm{vec}(B)$ for any matrices
$A,B,C$ \cite{Horn:1991tx}. Here $^*$ denotes complex conjugation and $^{\mathrm{T}}$ denotes transposition. The Schr\"odinger (von Neumann) equation describing evolution by a Hamiltonian $\hat{H}(t)$ translates to
\beq
\frac{\partial}{\partial t}\mathcal{E}(t) = -\frac{i}{\hbar} \left[ \hat{1}\otimes \hat{H}(t) - \hat{H}^{\mathrm{T}}(t) \otimes \hat{1} \right] \mathcal{E}(t) \equiv -\frac{i}{\hbar} \mathcal{H}(t) \mathcal{E}(t),
\eeq
with initial condition $\mathcal{E}(0) = \mathbf{I}$. In this appendix, we develop the generalized DHEOM that describs the reduced equations of motion for these propagators. 

Similar to the density matrix derivation, we begin by defining $\mathcal{E}(t,\Omega) \equiv \mathcal{E}_\Omega(t)P(\Omega,t)$ where $\mathcal{E}_\Omega$ is the conditioned propagator $\mathcal{E}_\Omega(t) \equiv \mathcal{E}(t | \Omega(t))$. This quantity satisfies a propagator version of the stochastic Liouville equation:
\beq
\frac{\partial \mathcal{E}(t, \Omega)}{\partial t} = \left[
  -\frac{i}{\hbar}\mathcal{H}_0 - \frac{i}{h}\Omega(t)\mathcal{V}
  \right] \mathcal{E}(t,\Omega) + \Gamma_\Omega\mathcal{E}(t,\Omega),
\eeq
where we have split the Hamiltonian into a time-independent component and a Markov fluctuating component: $\mathcal{H} = \mathcal{H}_0 + \Omega(t) \mathcal{V}$. As with the density matrix, the averaged propagator is given by the marginal:
\beq
\mathcal{E}(t) = \int_\mathcal{D} d\Omega~ \mathcal{E}(t,\Omega).
\eeq
When $\Omega(t)$ is a diffusive process with a generator possessing
polynomial eigenfunctions, we expand $P(\Omega,t)$ in generator
eigenfunctions $f_n(\Omega)$ as
\beq
\mathcal{E}(t, \Omega) \equiv \mathcal{E}_\Omega(t)P(\Omega, t) = \sum_{n=0}^\infty \bar{\mathcal{E}}_n(t) f_n(\Omega).
\eeq
Then utilizing the three-term recurrence relations that the orthogonal polynomial eigenfunctions follow, we arrive at the following classical hierarchical equations of motion for the propagator:
\bqa
\frac{\partial}{\partial t}\bar{\mathcal{E}}_0(t) &=& -\left[ \frac{i}{\hbar}\left( \mathcal{H}_0 -\frac{B_0}{A_0} \mathcal{V} \right) + \lambda_0 \right]  \bar{\mathcal{E}}_0(t)  -\frac{i}{\hbar} \frac{C_1}{A_1}\ \mathcal{V} \bar{\mathcal{E}}_{1}(t), \nn \\
\frac{\partial}{\partial t}\bar{\mathcal{E}}_n(t) &=& -\left[ \frac{i}{\hbar}\left( \mathcal{H}_0 -\frac{B_n}{A_n} \mathcal{V} \right) + \lambda_n \right] \bar{\mathcal{E}}_n(t) -\frac{i}{\hbar} \frac{C_{n+1}}{A_{n+1}}\ \mathcal{V} \bar{\mathcal{E}}_{n+1}(t) - \frac{i}{\hbar}\frac{1}{A_{n-1}}  \mathcal{V} \bar{\mathcal{E}}_{n-1}(t), ~~~~~~~ n > 0,
\label{eq:prop_hierarchy}
\eqa
where $\lambda_n, A_n, B_n$ and $C_n$ are eigenvalue and eigenfunction recurrence relation coefficients  defined identically as in the main text. The terminator equation for this DHEOM is explicitly:
\beq
\frac{\partial}{\partial t}\bar{\mathcal{E}}_N(t) = -\left[ \frac{i}{\hbar}\left( \mathcal{H}_0 -\frac{B_N}{A_N} \mathcal{V} \right) + \lambda_N \right] \bar{\mathcal{E}}_N(t) - \frac{1}{\hbar^2} \frac{C_{N+1}}{\lambda_{N+1} A_{N+1}A_N} ~\mathcal{V} \mathcal{V}~\bar{\mathcal{E}}_{N}(t) - \frac{i}{\hbar}\frac{1}{A_{N-1}} ~ \mathcal{V}~ \bar{\mathcal{E}}_{N-1}(t).
\eeq

\end{widetext}

\end{document}